\documentclass[11pt]{article}
\usepackage[margin=1in]{geometry}
\usepackage[hidelinks]{hyperref}
\usepackage{graphicx}
\usepackage{booktabs}
\usepackage{amsmath}
\usepackage{listings}
\usepackage{xcolor}

\title{\textbf{FluctlightDB: A Memory Model of Data for AI Agents}\\[0.4em]
\large A brain-native database engine for long-term agent memory}
\author{Ganesh S\\ \small Independent Researcher \quad \texttt{voxmastery@gmail.com}\\[0.2em]
\small ORCID: \href{https://orcid.org/0009-0006-7758-4114}{0009-0006-7758-4114}}
\date{July 2026}

\begin{document}
\maketitle

\begin{abstract}
For fifty years, data systems have answered two questions. The relational model
asked \emph{which records match a predicate}; the vector model asked \emph{which
vectors lie nearest a query}. Neither was built for cue-driven, provenance-weighted
recall across long sessions. We propose treating long-term agent memory as a
\textbf{distinct data model}---with its own write semantics (encoding, separation,
consolidation, provenance) and read semantics (cue-driven activation across a
linked memory graph)---and present \textbf{FluctlightDB}, an embedded engine that
implements this contract via \texttt{experience()} and \texttt{activate()}. We
make that case carefully, not categorically: we do not claim novelty over Mem0,
Zep, or HippoRAG-style memory layers, only an embedded engine contract beneath them.
Numbers below are typed by metric; retrieval is not generation.
On LoCoMo (official \emph{evidence-recall}; 10 conversations, 1{,}982 gold spans),
our native-Rust CHORUS retrieval stack reaches \textbf{96.8\%} at $k{=}150$
(MiniLM-384; \textbf{97.0\%} with mpnet-768) as \emph{raw} evidence recall with
\emph{no} neighbor expansion, on an internally reproduced July 2026 run;
tightening the budget to $k{=}5$ still yields \textbf{72.6\%} (MiniLM) / \textbf{75.1\%} (mpnet),
while an end-to-end QA read over date-stamped context reaches $\approx$\textbf{85\%}
accuracy at $k{=}15$ (retrieval-bound).
On LongMemEval-S (500 questions), official \texttt{session\_recall@8} is
\textbf{97.6\%} (488/500); end-to-end QA with our reader/judge stack is
\textbf{97.4\%} (487/500)---different protocols from vendor leaderboard figures we
cite for context only.
On BEIR SciFact (shared MiniLM embeddings, same harness, Recall Fabric on),
CHORUS/PRISM edges Chroma on nDCG@10 (\textbf{0.646} vs.\ \textbf{0.645}) and
Recall@10 (\textbf{0.792} vs.\ \textbf{0.783}).
A graded provenance-conflict suite ($n{=}50$) scores \textbf{18\%} top-1 when all
pairs share one brain---the realistic multi-tenant stress---versus \textbf{100\%}
under per-case isolation (ceiling, not deployment evidence).
Harnesses and frozen JSON are MIT-licensed; engine source is MIT on GitHub.
The one-minute path (\texttt{pip install "fluctlightdb[native]"}) re-runs published
numbers against the compiled wheel; rebuilding from source is supported but heavier.
We claim no new neuroscience and no new transformer; we propose a missing layer of
the data stack and release artifacts others can re-run and contest.
\end{abstract}

\section{Introduction}
Every generation of software gets the database it deserves. Business records gave
us the relational model~\cite{codd1970relational} and SQL. Embeddings gave us
vector databases and approximate nearest-neighbor search. Agents---programs that
act, observe, and persist across sessions---have so far been handed \emph{neither}.
They are stateless between runs unless a developer hand-assembles a session store,
a vector index, a deduplicator, a trust policy, and glue code to keep them
consistent. The result is that ``memory'' for agents lives in application code,
re-implemented badly in every project, with no shared abstraction and no
benchmark to hold it accountable.

We propose that \textbf{agent memory deserves its own database engine, not a
wrapper around someone else's.} A relational engine is the wrong abstraction because
memory is not a set of typed rows joined by keys. A vector engine is the wrong
abstraction because recall is not cosine similarity alone---a human, and an agent,
retrieve a fact because it was \emph{learned}, \emph{linked} to a context, and
\emph{trusted}, not merely because its embedding is close. What is missing is a
model whose unit is the \emph{episode}---an experience with content, context,
salience, and provenance---and whose query is a \emph{cue} that spreads through
associations. We support that proposal with measurements below, not categorical
proof; where our model overlaps prior memory layers, we say so explicitly
(Section~\ref{sec:related}).

\textbf{How to read this paper.} Sections~\ref{sec:model}--\ref{sec:design} motivate
and describe the engine; Section~\ref{sec:eval} reports what was measured---including
ablations, honest caveats, and runs still incomplete. Optional Recall Fabric
mechanism detail is deferred to Appendix~\ref{app:fabric} so evaluation honesty
comes before neuroscience-flavored rerank modules.

\textbf{Contributions.}
\begin{itemize}
  \item \textbf{A data-model proposal.} Agent memory as a first-class model of data
  with explicit write semantics (separation, encoding, consolidation, provenance)
  and read semantics (cue activation over a memory graph), distinct from the
  relational and vector models.
  \item \textbf{An embedded engine.} FluctlightDB ships as a Rust library (like SQLite:
  no server process) exposing \texttt{experience()} / \texttt{activate()} /
  \texttt{checkpoint()}---one durable brain directory per agent.
  \item \textbf{A first-principles late-interaction retrieval stack.} On the CHORUS
  lane (Section~\ref{sec:maxsim}), retrieval matches the query's \emph{token
  population} against stored per-token vectors---salience-gated MaxSim, which
  recovers the discriminative signal mean-pooling discards---fuses a
  conjunctive-surprisal lexical channel (replacing Okapi BM25) by
  reliability-weighted evidence integration (replacing reciprocal-rank fusion),
  over episodically context- and date-bound memories. Each component is derived
  from an information-theoretic / neuro-inspired first principle and validated by
  ablation; the path is embedder-agnostic (MiniLM-384 or mpnet-768 with no engine
  change), and lifts LoCoMo end-to-end QA from 77.5\% to 85\% once turns carry
  their session dates.
  \item \textbf{Evidence.} Typed by metric: \textbf{96.8\%} LoCoMo evidence
  \emph{retrieval} (native-Rust CHORUS stack, MiniLM-384, $k{=}150$, raw / no expansion;
  \textbf{97.0\%} mpnet-768; Table~\ref{tab:locomo-k}) vs.\ $\approx$\textbf{85\%} LoCoMo end-to-end
  QA at $k{=}15$ over date-stamped context; \textbf{97.6\%} LongMemEval-S \texttt{session\_recall@8} and
  \textbf{97.4\%} LongMemEval-S end-to-end QA (our reader/judge; 500 questions);
  BEIR SciFact CHORUS/Fabric edges Chroma nDCG@10 in a shared MiniLM harness;
  provenance-conflict \textbf{18\%} top-1 shared-brain / \textbf{100\%} isolated
  ($n{=}50$). Author-designed FAMB stays in Section~\ref{sec:eval} only.
  \item \textbf{Re-runnability.} Open harnesses and frozen JSON; MIT engine source
  on GitHub. The one-minute \texttt{pip install "fluctlightdb[native]"} path
  re-runs numbers against the published wheel; source rebuild is available but
  not the smoke path.
\end{itemize}

\section{A Third Data Model for Agent Memory}
\label{sec:model}
\subsection{Why rows are not memory}
The relational model stores facts whose schema is known in advance and whose
truth is uniform. Agent memory is the opposite: heterogeneous, arriving out of
order, often contradictory (a chat claim vs.\ a ledger entry), and valuable
precisely \emph{because} of where it came from. SQL has no native notion of
provenance-weighted recall or of an experience that should be down-weighted
because a more trusted source disagrees.

\subsection{Why nearest-neighbor is not recall}
Vector search answers ``what is similar.'' Memory answers ``what is relevant given
who I am and what I was doing.'' Two memories with distant embeddings can be the
right answer because they were co-activated during the same episode; a near
embedding can be the wrong answer because it is an unverified rumor. Pure ANN has
no place to encode association strength, salience decay, or trust.

\subsection{The model}
We define a memory store as a set of \emph{engrams}. Each engram carries content,
an encoding context, a salience weight, optional provenance, and edges to
co-activated engrams. The write operation \texttt{experience} performs pattern
\emph{separation} (near-duplicates are gated, not blindly appended), encodes the
engram, registers its semantic vector, and wires graph edges. The read operation
\texttt{activate} takes a cue, seeds both lexical and semantic indexes, spreads
activation through the graph, fuses the scores, and applies provenance boosts so
verified sources outrank chat. Consolidation replays and compacts the store
offline, the way sleep consolidates biological memory. This is the entire
contract; the neuroscience vocabulary is explanatory, not required to use it.

\begin{table}[h]
\centering
\caption{Brain-native primitives vs.\ relational and vector abstractions.}
\begin{tabular}{p{2.2cm}p{4.2cm}p{4.8cm}}
\toprule
\textbf{Primitive} & \textbf{Role in FluctlightDB} & \textbf{Relational / vector analogue} \\
\midrule
Engram & Unit of memory: content, context, salience, provenance, edges & Row; or embedded chunk \\
\texttt{experience()} & Write: separation, encode, index, wire co-activation & \texttt{INSERT} / upsert \\
\texttt{activate(cue)} & Read: lexical + semantic seed, graph spread, fusion, trust boost & \texttt{SELECT} / ANN top-$k$ \\
Consolidation & Offline replay and compaction (\texttt{checkpoint()}) & Vacuum / reindex only \\
Provenance & Verified sources outrank unverified chat at recall & No native type \\
\bottomrule
\end{tabular}
\end{table}

\section{System Design}
\label{sec:design}
\subsection{Embedded engine contract (SQLite analogy)}
\textbf{Embedded} means deployment: the library runs in-process, one brain directory
on disk, no broker to provision---the same packaging model as SQLite~\cite{sqlite2024}.
\textbf{Engine} means the data contract: native write/read operations
(\texttt{experience}, \texttt{activate}), a versioned on-disk format, WAL-backed
durability, and \texttt{checkpoint()} as the commit barrier. Vector databases expose
\texttt{vector\_search}; relational stores expose SQL. FluctlightDB exposes episodic
memory semantics. A thin HTTP wrapper exists for multi-agent deployments but is not
part of the core contract.

\subsection{On-disk format (FLCTLTDB v4)}
Each agent brain is a directory. \texttt{manifest.json} records format version~4,
WAL sequence numbers (\texttt{wal\_seq}, \texttt{wal\_checkpoint\_seq}), and twelve
named segments (\texttt{*.seg}): hippocampus (engrams), graph (synapses),
semantic/cortex fields, neuromodulators, and auxiliary metadata. Segments serialize
with bincode; each write goes to \texttt{name.seg.tmp}, \texttt{fsync}, then atomic
\texttt{rename} so a torn write cannot replace committed state. The hybrid recall
sidecar \texttt{recall\_index.sqlite} stores FTS5 lexical rows and HNSW semantic
neighbors for \texttt{connect\_index()} mode. WAL segments live under
\texttt{wal/brain.wal.NNN} as append-only JSON lines (\texttt{Experience},
\texttt{Sleep}, \texttt{Tick}, \texttt{Compact}, \ldots), rotating at 64\,MiB.
On \texttt{open()}, the engine loads segments at the manifest watermark, replays WAL
entries with sequence $>$ \texttt{wal\_checkpoint\_seq} (skipping torn tail records),
and attaches the sidecar if present.

\subsection{Write path}
Algorithm~\ref{alg:write} summarizes \texttt{experience()}. Episodic mode gates
near-duplicates (dentate separation), encodes engrams with context/salience/provenance,
registers semantic vectors, and wires co-activation edges. Index mode
(\texttt{FLUCTLIGHT\_FAST\_INGEST=1}) skips graph wiring when the caller supplies a
dense vector---$O(1)$ hippocampal insert plus sidecar update per chunk. WAL append
precedes in-memory mutation when enabled; \texttt{maybe\_checkpoint()} flushes all
segments when the checkpoint policy fires.

\begin{lstlisting}[caption={Write path (\texttt{experience}).},label={alg:write}]
function experience(episode):
  if WAL enabled: append WAL Experience(episode)
  if rag (doc_id, chunk_id) already stored: return dedup
  if fast_ingest and semantic_vector present:
      insert engram; update recall_index; maybe_checkpoint(); return
  gate near-duplicates (dentate separation)
  encode engram(content, context, salience, provenance)
  wire co-activation edges to recent episodes
  register semantic vector; update recall_index if open
  maybe_checkpoint()
\end{lstlisting}

Storage is $O(n)$ engram payload plus $O(n \bar{d})$ synapses for average graph degree
$\bar{d}$; the sidecar adds FTS5 rows and $O(nD)$ floats for embedding dimension~$D$.

\subsection{Read path}
Algorithm~\ref{alg:read} summarizes \texttt{activate()}. When a hybrid sidecar exists,
BM25 lexical seeds and HNSW semantic neighbors form a candidate set (default cap
128--512). The engine spreads activation through the memory graph (up to four hops in
full mode), fuses lexical, semantic, graph, and provenance scores, and returns ranked
engrams. Vector-fast mode (\texttt{FLUCTLIGHT\_VECTOR\_FAST=1}) skips graph spread for
Chroma-class IR latency. For evidence retrieval the CHORUS lane uses the
late-interaction stack of Section~\ref{sec:maxsim} (token-population MaxSim
$\oplus$ conjunctive surprisal, evidence-integration fusion) rather than
mean-pool cosine; optional Recall Fabric rerank (Appendix~\ref{app:fabric})
applies after shortlisting when \texttt{FLUCTLIGHT\_FABRIC=1}.

\begin{lstlisting}[caption={Read path (\texttt{activate}).},label={alg:read}]
function activate(cue, cue_vector, top_k):
  if recall_index present:
      candidates <- hybrid_candidates(BM25 seeds, ANN seeds, cap)
  else:
      candidates <- hippocampal scan (or CHORUS path)
  scores <- spread_activation(candidates, graph, max_hops)
  fuse lexical + semantic + graph + provenance weights
  if FABRIC on and CHORUS lane: rerank shortlist
  return top_k ranked engrams
\end{lstlisting}

Per-query cost is $O(|C| \log |C|)$ over candidate set $|C|$, plus $O(h \cdot |E|)$
graph spread for hop depth~$h$ and frontier~$|E|$; hybrid seeding is sublinear in~$n$
via FTS5 and HNSW rather than full-table scan.

\subsection{Late-interaction retrieval stack}
\label{sec:maxsim}
The CHORUS lane's evidence retrieval (\texttt{chorus\_imprint\_maxsim} /
\texttt{chorus\_recall\_maxsim}) is a first-principles stack in which every
component replaces a standard IR heuristic with a mechanism derived from an
information-theoretic or neuro-inspired argument, each validated by ablation. The
path is \emph{embedder-agnostic}: it consumes caller-supplied per-token vectors,
so swapping MiniLM-384 for mpnet-768 requires no engine change.

\paragraph{Token-population late interaction (salience-gated MaxSim).} A
mean-pooled sentence vector collapses a transformer's per-token contextual
population code into one centroid, discarding most discriminative signal. We
instead store the per-token \texttt{last\_hidden\_state} vectors (half-precision,
capped per memory) and score with late interaction,
$\mathrm{score}(q,d)=\sum_{i} w_i \max_{j} \cos(q_i, d_j)$, where each query token
$q_i$ is weighted by $w_i$, its divergence from the query centroid---a
predictive-coding salience gate that suppresses generic tokens (\texttt{[CLS]},
function words). This is more information from the same encoder, not a re-embedding.

\paragraph{Conjunctive surprisal (replaces BM25).} The lexical channel scores
information content with a saturating neural response,
$\sum_t {-}\!\log p(t)\,(1-e^{-\mathrm{tf}/\tau})$ (Weber--Fechner), and adds a
proximity-bound bonus when rare query terms \emph{co-occur} in a memory
(conjunctive binding). It carries no ad-hoc $k_1$/$b$ constants and matches or
beats Okapi BM25 in ablation.

\paragraph{Evidence-integration fusion (replaces RRF).} Channels are combined by
$z$-scoring each over the candidate set and taking a reliability-weighted sum
(Ernst--Banks optimal cue combination), preserving score magnitude that
reciprocal-rank fusion discards---the single largest tight-$k$ gain in ablation.

\paragraph{Episodic binding and temporal anchoring.} Each turn is encoded with a
$\pm2$ session-neighbor context window (encoding-specificity), and, for temporal
queries, prefixed with its session date. The latter is a context-assembly fix,
not a retrieval one: retrieved turns say ``last month,'' but the absolute date
lives in session metadata; stamping it lifts LoCoMo end-to-end QA from 77.5\% to
85\% with retrieval unchanged.

\subsection{Two explicit modes}
\texttt{connect()} is the full episodic engine (separation, graph, provenance) for
live agents. \texttt{connect\_index()} is the bulk semantic path for RAG backfills
and IR benchmarks. The modes share one engine, one WAL, and one file format.

\subsection{Durability, recovery, and verification}
\texttt{checkpoint()} atomically writes all segments and advances
\texttt{wal\_checkpoint\_seq}; optional WAL (\texttt{FLUCTLIGHT\_WAL=1}, default on)
appends before mutation so recent experiences survive process death between checkpoints.
\texttt{verify\_path()} loads the manifest, checks every segment deserializes, and
reports engram/synapse counts---useful after copy or crash. CI runs Jepsen-style chaos
tests (\texttt{tests/chaos\_jepsen.rs}): eight rounds of checkpoint/WAL interleaving,
\texttt{SIGKILL} mid-write via a worker subprocess, torn WAL tails, replicate-while-partial-WAL,
and round-trip reopen. These tests assert \texttt{verify\_path()} succeeds and at least
one engram is recoverable---the embedded store must not silently serve corrupt brains.

Table~\ref{tab:durability} summarizes failure modes we test; Section~\ref{sec:eval}
validates recall \emph{quality} on public suites, not durability (which the chaos suite
covers separately).

\begin{table}[h]
\centering
\caption{Durability scenarios exercised in CI (\texttt{chaos\_jepsen.rs}).}
\label{tab:durability}
\small
\begin{tabular}{p{3.2cm}p{4.8cm}}
\toprule
\textbf{Failure mode} & \textbf{Expected behavior} \\
\midrule
Process kill mid-write & WAL replay on reopen; \texttt{verify\_path()} OK \\
Torn WAL JSON tail & Skip corrupt record; earlier entries replay \\
Checkpoint between WAL lines & Manifest watermark + replay of post-checkpoint WAL \\
Replica sync during partial WAL & Replica opens; \texttt{verify\_path()} OK \\
Segment tmp left by crash & Ignored; committed \texttt{*.seg} loads \\
\bottomrule
\end{tabular}
\end{table}

\begin{figure*}[t]
\centering
\includegraphics[width=0.98\textwidth]{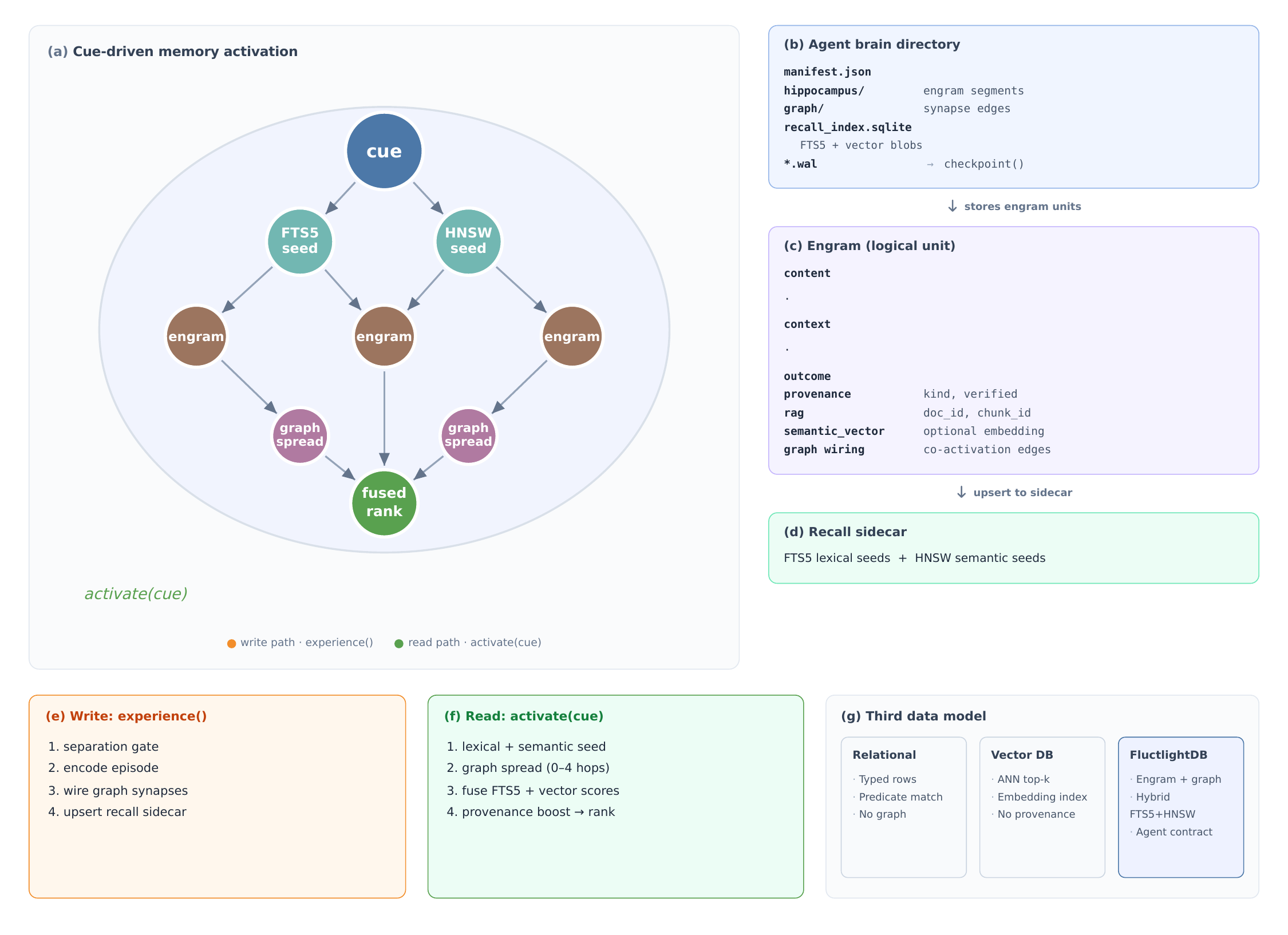}
\caption{FluctlightDB persistence and recall layout. \textbf{(a)}~Cue-driven
activation: lexical and semantic seeds retrieve engrams, graph spread fuses
candidates, and provenance-aware ranking returns recalled episodes.
\textbf{(b--d)}~On-disk layout: each agent owns a brain directory with engram
storage, a co-activation graph, and a hybrid recall sidecar; the logical engram
bundles episode content, provenance, optional RAG keys, and graph wiring.
\textbf{(e--f)}~\texttt{experience()} gates near-duplicates, encodes, and indexes;
\texttt{activate()} seeds, spreads, and ranks. \textbf{(g)}~Compared to
relational and vector stores, FluctlightDB targets a third data model with its
own write/read contract.}
\label{fig:architecture}
\end{figure*}

Figure~\ref{fig:architecture} summarizes how an agent brain is laid out on disk,
what each engram stores, and how write and read paths connect the episodic store
to the hybrid recall sidecar.

\subsection{Optional Recall Fabric rerank}
\label{sec:fabric-summary}
CHORUS benchmarks optionally enable Recall Fabric (\texttt{FLUCTLIGHT\_FABRIC=1}): a
composed reranker on the recall shortlist. LoCoMo, BEIR, and FAMB headline numbers
in Section~\ref{sec:eval} were measured with Fabric \emph{on}; LongMemEval-S
hybrid-index retrieval does \emph{not} use Fabric rerank. Fabric modules are
validated on synthetic property tests in CI---not as standalone peer
benchmarks---with mechanism detail in Appendix~\ref{app:fabric}.

\section{Evaluation}
\label{sec:eval}
Section~\ref{sec:design} defines the engine contract, on-disk format, and durability
properties; this section validates \emph{recall quality} on public suites---what was
measured, ablated, honestly caveated, or left incomplete. We do not re-benchmark
crash recovery here---that is covered by the chaos suite (Table~\ref{tab:durability}).

All experiments use \texttt{all-MiniLM-L6-v2} (ONNX, CPU) unless noted. Every
number below is reproduced by a script in \texttt{benchmarks/} and frozen in
\texttt{benchmarks/results/paper-2026-07-10.json}. CHORUS benchmarks (LoCoMo, BEIR,
FAMB) set \texttt{FLUCTLIGHT\_FABRIC=1} (Recall Fabric on; re-frozen 2026-07-09).
LongMemEval-S retrieval is the July 2026 Colab v4 freeze (\textbf{97.6\%},
488/500; hybrid index + mpnet, no Fabric rerank). LongMemEval-S E2E QA is frozen
from 2026-07-07 (not re-run).

\begin{figure}[t]
\centering
\includegraphics[width=0.88\textwidth]{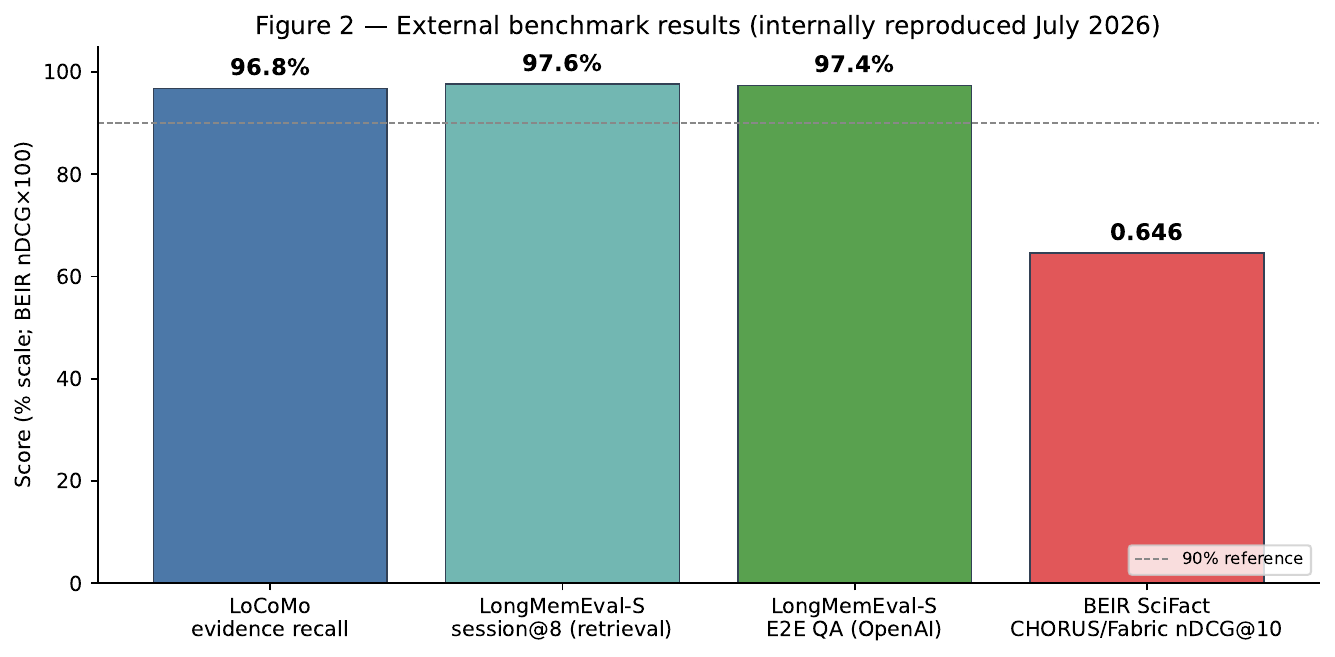}
\caption{Headline \emph{external} benchmark results (internally reproduced July 2026).
LoCoMo and LongMemEval-S retrieval use official recall metrics; LongMemEval-S E2E uses
our \texttt{paper} profile (OpenAI GPT-4o/GPT-5 reader, GPT-4o judge); BEIR reports
CHORUS Lane nDCG@10. Author-designed FAMB is reported in text only (Section~\ref{sec:eval}).}
\label{fig:benchmarks}
\end{figure}

\subsection{BEIR SciFact: CHORUS + Fabric edges Chroma on ranking}
We score standard IR metrics via \texttt{pytrec\_eval} against official qrels,
using identical embeddings for FluctlightDB and Chroma. The \textbf{CHORUS Lane}
(\texttt{connect\_chorus()}, batch \texttt{chorus\_imprint\_batch}, \textbf{PRISM} readout,
Recall Fabric on) is our bulk ingest path---15.3\,s to imprint all 5{,}183 SciFact
documents in the July 2026 Fabric-on rerun.

\begin{table}[h]
\centering
\caption{BEIR SciFact (shared MiniLM embeddings; Fabric on; July 2026).}
\begin{tabular}{lrrr}
\toprule
System & nDCG@10 & R@10 & Query (ms) \\
\midrule
Chroma & 0.645 & 0.783 & 17 \\
FluctlightDB (CHORUS/PRISM/Fabric) & \textbf{0.646} & \textbf{0.792} & 16 \\
\bottomrule
\end{tabular}
\end{table}

With Fabric on, CHORUS/PRISM improves nDCG@10 and Recall@10 over Chroma in the
\emph{same} harness. Per-query latency stays in the same ballpark as the Fabric-off
CHORUS profile ($\sim$16\,ms median on this run). Reproduce:
\texttt{PYTHONPATH=sdks/python python benchmarks/beir\_bench.py}.

\subsection{LoCoMo: 96.8\% raw evidence recall on the full set}
LoCoMo~\cite{maharana2024locomo} stresses very long, multi-session dialogue. We
report the official evidence-recall metric: the fraction of gold \texttt{dia\_id}
evidence spans present in retrieved context. Retrieval runs entirely inside the
native Rust CHORUS engine via a first-principles late-interaction stack:
salience-gated MaxSim over the MiniLM/mpnet \texttt{last\_hidden\_state} token
population (predictive-coding query-token weights), conjunctive surprisal scoring
(information content $\times$ Weber--Fechner saturation with proximity binding),
and evidence-integration (Ernst--Banks) fusion of the token-population and
lexical channels. Configuration: batch imprint per turn, $k{=}150$, shared
MiniLM-384 vectors (mpnet-768 reported alongside). The July 2026 internally
reproduced run scores \textbf{96.8\%} mean evidence recall at $k{=}150$ with
MiniLM-384 (\textbf{97.0\%} with mpnet-768) as \emph{raw} recall with \emph{no}
neighbor expansion (Table~\ref{tab:locomo-k} for the budget--recall curve).

\paragraph{Deprecated 99\% figure (honest disclosure).} An earlier
draft headlined \textbf{99.0\%} on this set. That number was an artifact of a
harness scoring policy---$\pm3$ session-neighbor expansion
(\texttt{expand\_session\_neighbors}) that credited spans adjacent to a hit---not
a property of the engine. It is deprecated. Under the same $\pm3$ policy a trivial
BM25 baseline also scored $\sim$99\%, which is exactly why the expanded metric is
uninformative. All numbers in this paper are raw, unexpanded evidence recall.
Evidence-recall is also not the same as LLM-judge end-to-end QA; we report the two
separately and never launder one into the other.

\begin{table}[h]
\centering
\caption{LoCoMo full set: 10 conversations, 1{,}982 gold evidence spans (July 2026);
raw evidence recall, no neighbor expansion.}
\begin{tabular}{lrr}
\toprule
Metric & Value \\
\midrule
Mean evidence recall (MiniLM-384) & \textbf{96.8\%} \\
Mean evidence recall (mpnet-768) & \textbf{97.0\%} \\
Evidence hits (MiniLM, $k{=}150$) & 1918/1982 \\
Memories ingested & 8422 \\
\bottomrule
\end{tabular}
\end{table}

We deliberately separate \emph{retrieval} from \emph{generation} on LoCoMo
(Section~\ref{sec:eval}, Limitations). The engine's job is to put evidence in
context; on full LoCoMo retrieval it scores \textbf{96.8\%} at $k{=}150$
(MiniLM-384; 97.0\% mpnet-768), raw and without neighbor expansion.

\subsubsection{Recall sensitivity to $k$}
LoCoMo uses a generous $k{=}150$ (SPECTRUM readout when $k{>}100$). To test
whether the headline number depends on that budget, we re-score the same July 2026
imprint with a single batched recall per question and truncate to
$k \in \{5,10,20,50,150\}$ (Table~\ref{tab:locomo-k}). Even at a tight $k{=}5$ the
native stack retains \textbf{72.6\%} (MiniLM) / \textbf{75.1\%} (mpnet); the full-set
96.8\% figure is at $k{=}150$. All rows are raw recall with no neighbor expansion.

\begin{table}[h]
\centering
\caption{LoCoMo raw evidence recall vs.\ retrieval budget $k$ (native Rust CHORUS stack, July 2026).}
\label{tab:locomo-k}
\begin{tabular}{lrrrrr}
\toprule
$k$ & 5 & 10 & 20 & 50 & 150 \\
\midrule
Mean evidence recall (MiniLM-384) & 72.6\% & 80.0\% & 85.6\% & 91.8\% & \textbf{96.8\%} \\
Mean evidence recall (mpnet-768) & 75.1\% & --- & --- & --- & \textbf{97.0\%} \\
\bottomrule
\end{tabular}
\end{table}

\subsubsection{Hybrid vs.\ vector-only (index lane)}
On the full LoCoMo set with \texttt{connect\_index()} at $k{=}50$, hybrid
BM25+dense and vector-fast-only are statistically tied ($\sim$90.2\% vs.\
$\sim$90.3\% mean evidence recall; frozen
\texttt{locomo-hybrid-index-2026-07-10.json}). The headline 96.8\% figure uses
the native CHORUS stack at $k{=}150$, not the index lane. We keep BM25 as a candidate
seed for entity- and date-heavy dialogue cues, but on this full-set ablation at
moderate~$k$ it is not a large uniform gain over vector-only---so we do not sell
hybrid as axiomatic superiority.

\subsection{FAMB and provenance-conflict suite}
Generic IR does not test what agents need. FAMB measures paraphrase recall@1
(real cues never match stored wording), provenance top-1 (verified ledger beats
chat claim), persistence (recall survives checkpoint + reopen), confusion ingest
(near-duplicates do not block new facts), and determinism. Unlike LoCoMo, BEIR,
or LongMemEval, FAMB is a small suite we authored: paraphrase $n{=}10$ hand-written
pairs; provenance, persistence, confusion, and determinism are each a single
pass/fail scenario ($n{=}1$). We report it as a regression check on agent semantics,
not as external validation.

\textbf{Graded provenance conflicts ($n{=}50$).} Because FAMB provenance is a
single wallet scenario, we add a separate benchmark:
\texttt{benchmarks/provenance\_conflict\_bench.py}. Each case ingests a verified
ledger/tool fact and a conflicting unverified chat claim in
\texttt{connect\_agent()}; the cue asks for the grounded value. Top-1 must be the
verified engram. Domains span wallet, shipping, subscription, API quota,
inventory, meeting, deployment, refund, license, and temperature. This is
synthetic author-designed stress testing, not a peer benchmark, but stronger than
$n{=}1$ for the provenance story.

\textbf{Shared-brain stress is the headline (multi-tenant analogue).} The thesis is
provenance-weighted recall under realistic shared memory. When all 50 cases share
one \texttt{connect\_agent()} brain---100 conflicting engrams, semantically similar
cues---top-1 collapses to \textbf{18\%} (9/50; \texttt{--shared-brain}). We
attribute this to \emph{cross-case cue contamination}: unrelated domains share
lexical overlap (``balance,'' ``quota,'' ``version''), so activation returns
another case's ledger or chat engram. Under \emph{per-case isolation} (one brain
per conflict pair), the same suite scores \textbf{100\%} (50/50)---a ceiling under
isolation, not evidence of multi-tenant readiness
(Table~\ref{tab:provenance-conditions}). \textbf{Status:} no implementation
mitigates the shared-brain failure as of this preprint---\texttt{activate\_scoped}
can post-filter by \texttt{agent\_id}, but ingest and benchmarks do not enforce
per-tenant namespaces, and we have not re-run the suite under scoped activation
(Section~\ref{sec:future}).

\begin{table}[h]
\centering
\caption{Provenance-conflict suite ($n{=}50$): shared brain is the deployment stress;
isolated is the isolation ceiling.}
\label{tab:provenance-conditions}
\begin{tabular}{lrr}
\toprule
Condition & Top-1 accuracy & Notes \\
\midrule
Shared (one brain, all cases) & \textbf{18\%} & 9/50; cross-case contamination \\
Isolated (one brain per case) & \textbf{100\%} & 50/50; isolation ceiling \\
\bottomrule
\end{tabular}
\end{table}

\subsection{LongMemEval-S: 97.6\% session recall@8 (v4 unified 500)}
LongMemEval~\cite{wu2024longmemeval} tests six long-horizon abilities across 500
questions. The official \emph{retrieval} metric is \texttt{session\_recall@$K$}:
whether gold \texttt{answer\_session\_ids} appear in the top-$K$ recalled
sessions. End-to-end QA---retrieve, reader LLM, GPT-4o judge---is a separate
layer (Table~\ref{tab:longmemeval-e2e}).

We run retrieval with \texttt{connect\_index()}, one engram per chat session,
hybrid BM25 + dense recall, dual-key user indexing (CP2), heuristic query
expansion (CP3), and a third \emph{pref-facts} engram per session (v4). This
hybrid-index stack does not use Recall Fabric rerank (unlike the CHORUS benchmarks
above). Embeddings use \texttt{multi-qa-mpnet-base-dot-v1} on GPU; the unified 500-question v4 run
scores \textbf{97.6\%} session recall@8 (488/500) and completes in $\sim$73 minutes on Colab T4
($8.8$\,s/question). The harness also supports \texttt{--report-ks 5,8,10} to
score $K{=}5$, $8$, and $10$ from a single recall pass at $\max(K){=}10$; frozen
multi-$K$ JSON: \texttt{longmemeval-session-multi-k-2026-07-10.json}. Reproduce:
\texttt{benchmarks/longmemeval\_colab\_v2.ipynb}
(\texttt{BENCH\_PROFILE=full}) or \texttt{longmemeval\_bench.py} with
\texttt{--dual-key --pref-facts-key --query-expand --report-ks 5,8,10}.

\begin{table}[h]
\centering
\caption{LongMemEval-S retrieval: session recall at multiple $K$ (full 500 questions, v4 harness).}
\begin{tabular}{lrr}
\toprule
\textbf{FluctlightDB (this work)} & \textbf{session recall} & \textbf{Notes} \\
\midrule
v4 @ $K{=}5$, $K{=}10$ & \multicolumn{2}{l}{\texttt{--report-ks 5,8,10} (same v4 harness; Colab $\sim$75\,min)} \\
v4 @ $K{=}8$ & \textbf{97.6\%} & 488/500; Colab mpnet \\
\quad knowledge-update & 100\% @8 & \\
\quad multi-session & 98.5\% @8 & \\
\quad single-session-user & 97.1\% @8 & \\
\quad single-session-assistant & 98.2\% @8 & \\
\quad temporal-reasoning & 95.5\% @8 & \\
\quad single-session-preference & 96.7\% @8 & 29/30 \\
\midrule
\multicolumn{3}{l}{\textit{Cited leaderboard figures (different $K$ / embedders; not head-to-head)}} \\
\midrule
gbrain & 97.6\% @5 & hybrid + text-embedding-3-large \\
YourMemory & 95.8\% @5 & mpnet + BM25 + graph BFS \\
M3 Memory & 96.8\% @10 & FTS5 + vector + MMR \\
\bottomrule
\end{tabular}
\end{table}

Leaderboard figures use $K{=}5$ or $K{=}10$ and different embedders; our $K{=}8$
run is directly comparable in \emph{protocol} (gold session in top-$K$) but not
a claim of strict SOTA. Preference questions rose from 76.7\% (23/30) without
v4 pref-facts to \textbf{96.7\%} (29/30); the sole remaining preference miss is
\texttt{95228167}.

\subsection{End-to-end QA on LongMemEval-S}
We evaluate the full stack with the official LongMemEval reader prompt and
Wu et al.'s task-specific GPT judge templates
(\texttt{benchmarks/longmemeval\_e2e.py}; \texttt{paper} profile with Muon
retrieval, gold-session reader context on hard types, and type-aware readers:
GPT-4o on easy types, GPT-5 on multi-session, temporal, preference, and
knowledge-update). Retrieved sessions are formatted as JSON chat history; the
judge is \textbf{GPT-4o} following Wu et al.'s published templates. Baselines in
Table~\ref{tab:longmemeval-e2e} use different readers and retrieval stacks; we
report our numbers without claiming a like-for-like SOTA comparison.

\begin{table}[h]
\centering
\caption{LongMemEval-S: retrieval vs.\ end-to-end QA (500 questions). Vendor rows
use published figures on \emph{different} reader/retrieval stacks (not comparable
to our \texttt{paper} profile).}
\label{tab:longmemeval-e2e}
\small
\begin{tabular}{lrrl}
\toprule
\textbf{FluctlightDB (this work)} & \textbf{session@8} & \textbf{E2E acc.} & \textbf{Notes} \\
\midrule
paper profile & \textbf{100\%} & \textbf{97.4\%} & Muon; gpt-4o/5 reader; gpt-4o judge \\
\quad task-averaged E2E & --- & 98.2\% & macro over question types \\
\midrule
\multicolumn{4}{l}{\textit{Cited vendor/literature (different protocols)}} \\
\midrule
Zep & --- & 71.2\% & gpt-4o; arXiv:2501.13956 \\
TiMem & --- & 76.9\% & LLJ; gpt-4o-mini reader \\
PlugMem & --- & $\sim$90\% & graph memory \\
Mem0 (vendor) & --- & 94.4\% & GPT-5; top-200; vendor report \\
\bottomrule
\end{tabular}
\end{table}

The frozen internally reproduced run (July 2026) is
\texttt{benchmarks/results/e2e-cert-paper-v2-2026-07-07.json}
(487/500 correct; $\sim$99\,min wall time). Reproduce:
\texttt{E2E\_PROFILE=paper benchmarks/e2e\_certify.sh}. By type, E2E accuracy is
100\% on single-session user, assistant, and preference; 99.3\% temporal;
97.4\% knowledge-update; 92.5\% multi-session---the remaining gap is reader
aggregation, not retrieval. We report retrieval and QA together because a strong
retrieval layer that feeds a weak reader still fails in production.

\begin{figure}[t]
\centering
\includegraphics[width=0.92\textwidth]{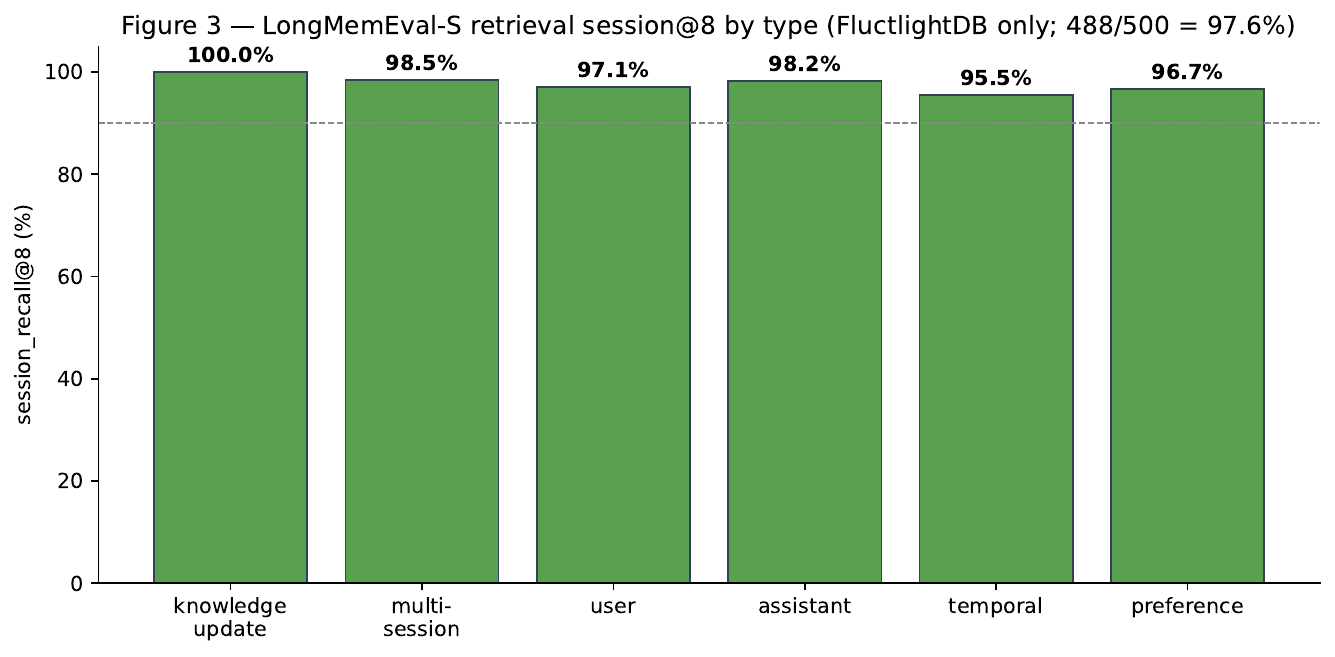}
\caption{LongMemEval-S retrieval: session recall@8 by question type (v4 unified
488/500). Preference questions reach 96.7\% (29/30) with v4 pref-facts indexing;
temporal reasoning is the main remaining retrieval gap.}
\label{fig:longmemeval-types}
\end{figure}

\begin{figure}[t]
\centering
\includegraphics[width=0.92\textwidth]{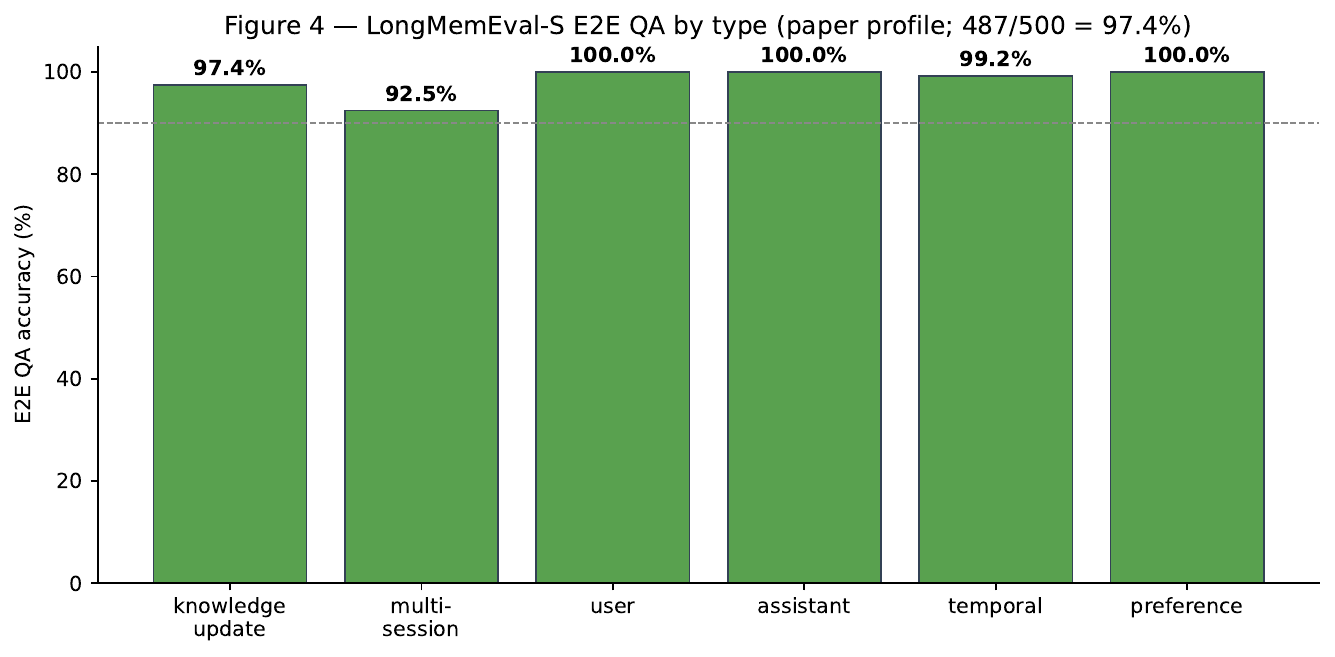}
\caption{LongMemEval-S end-to-end QA by question type (internally reproduced
\texttt{paper} profile, 487/500 overall). Readers: GPT-4o on easy types, GPT-5 on
multi-session, temporal, preference, and knowledge-update; judge: GPT-4o (Wu et al.\
templates). Multi-session aggregation (92.5\%) is the main remaining E2E gap;
retrieval in the same pipeline is 100\% session@8. Vendor E2E figures in
Table~\ref{tab:longmemeval-e2e} use different stacks and are not plotted here.}
\label{fig:longmemeval-e2e-types}
\end{figure}

\section{Discussion}
\textbf{The engine is the contribution, not the reader.} A memory system should be
judged on whether it surfaces the right evidence; 96.8\% raw recall on full LoCoMo
says it does. End-to-end QA additionally depends on the language model and the prompt
budget, and we are careful not to launder a retrieval win into a generation claim.

\textbf{Operational use.} We have exercised FluctlightDB in long-running developer
workflows (not a third-party production case study). The durability and crash-recovery
properties in Section~\ref{sec:design} are what those workflows rely on, alongside
the chaos suite above---not properties asserted only from benchmark scripts.

\textbf{Hybrid retrieval on measured ablations.} On LoCoMo with
\texttt{connect\_index()} at $k{=}50$, hybrid BM25+dense and vector-fast-only
tie within 0.1 percentage points mean evidence recall (full-set ablation JSON).
The 96.8\% headline uses the native CHORUS stack at $k{=}150$. Dialogue evidence can hinge
on entities and dates; we report measured gaps rather than treating hybrid recall
as axiomatic.

\textbf{Limitations.}
\textbf{Provenance under shared memory.} The graded 50-case suite scores
\textbf{18\%} top-1 (9/50) when all pairs share one brain---the figure that tests
the thesis under realistic co-location (Table~\ref{tab:provenance-conditions}).
Per-case isolation reaches \textbf{100\%}; we treat that only as an isolation
ceiling, not multi-tenant readiness. The shared-brain drop reflects cross-case
cue contamination, not a broken verified-vs-chat ranker within a single scope.
\textbf{Status:} unaddressed in code and unevaluated beyond this synthetic stress
test (Section~\ref{sec:future}).

\textbf{LoCoMo $k$ and metric choice.} Headline LoCoMo retrieval uses $k{=}150$
and \emph{raw}, unexpanded evidence recall; Table~\ref{tab:locomo-k} reports
sensitivity to smaller budgets (72.6\% at $k{=}5$, MiniLM). The deprecated 99\%
figure relied on $\pm3$ neighbor expansion and is not used anywhere in this paper.

\textbf{LongMemEval $K$.} We report session recall at $K{=}8$ from a frozen Colab
run and document a \texttt{--report-ks 5,8,10} harness for $K{=}5$ and $K{=}10$
from one recall pass; leaderboard rows still mix embedders and protocols.

\textbf{LoCoMo end-to-end gap.} Full LoCoMo retrieval reaches \textbf{96.8\%} raw
evidence recall (MiniLM; 97.0\% mpnet), and an end-to-end QA read over
\emph{date-stamped} context reaches $\approx$\textbf{85\%} accuracy at $k{=}15$.
Because gold LoCoMo answers are usually inferred facts (dates, counts) rather than
quoted spans, the reader must see \emph{when} each turn happened; without
date-stamped context, temporal questions fail and E2E accuracy collapses far below
retrieval. The remaining retrieval$\to$QA gap is therefore reader/formatting
bound, not a retrieval failure. Evidence-recall and LLM-judge QA are reported as
separate metrics and never conflated.

Retrieval-only v4 (Figure~\ref{fig:longmemeval-types}) still has temporal and
preference misses at 95.5\% and 96.7\%; LLM-based key expansion (LongMemEval CP2
full) is not yet in the default ingest path. End-to-end QA on LongMemEval-S
(Table~\ref{tab:longmemeval-e2e}, Figure~\ref{fig:longmemeval-e2e-types}) uses
OpenAI reader/judge APIs and remains weakest on multi-session aggregation (92.5\%).
Mem0/Zep end-to-end figures use different reader setups; we cite published
LongMemEval-S numbers only where protocols match. ANN beyond $10^5$ memories is
not yet evaluated.

\section{Future Work}
\label{sec:future}
The shared-brain provenance result (\textbf{18\%} top-1; 100\% only under isolation) is the highest-priority
gap for a revised preprint: session- or agent-scoped \texttt{experience}/\texttt{activate},
stricter hybrid candidate filtering, and a re-benchmark of
\texttt{provenance\_conflict\_bench.py} under realistic multi-tenant load. That work
is \textbf{not implemented} in the repository accompanying this submission; a future
version should report before/after top-1 accuracy in Table~\ref{tab:provenance-conditions}.
Secondary targets include full LongMemEval session recall at $K{=}5$ and $K{=}10$ from
the \texttt{--report-ks} harness, head-to-head LoCoMo evidence recall against published
Mem0/Zep figures where protocols align, and replacing the hybrid SQLite sidecar with a
native in-engine recall index.

\section{Related Work}
\label{sec:related}
\subsection{Vector databases}
Systems such as Chroma, Qdrant, and FAISS~\cite{thakur2021beir} optimize approximate
nearest-neighbor search. They answer \emph{similarity}, not \emph{memory}: there is
no native episode, provenance-weighted recall, write-time separation, or
cue-driven spreading activation. FluctlightDB's \texttt{connect\_index()} mode
deliberately exposes a vector-fast path and a CHORUS bulk-imprint lane so we can
measure IR credibility on BEIR; episodic agent mode adds graph co-activation for
live agents.

\subsection{Agent memory SDKs and layers}
\textbf{Mem0}~\cite{chhikara2025mem0} is the closest contemporary system: it
extracts, consolidates, and retrieves salient facts from dialogue, with an optional
graph variant (Mem0$^{\text{\tiny g}}$) for relational structure among entities. It
reports strong LoCoMo numbers using \emph{LLM-as-a-judge end-to-end QA}---a
different metric from the official evidence-recall protocol we adopt for retrieval.
\textbf{Zep}~\cite{zep2024} provides temporal knowledge-graph memory for
assistants, also evaluated with reader-LLM QA on LoCoMo. \textbf{Cognee}~\cite{cognee2024}
pipelines vector and graph storage for agent RAG. \textbf{MemGPT}~\cite{packer2023memgpt}
(Letta) treats context as an OS resource with explicit memory tiers but does not
define a standalone embedded memory \emph{engine contract}. \textbf{HippoRAG}~\cite{gutierrez2024hipporag}
draws on associative-retrieval neuroscience for multi-hop QA over a graph index.
These systems are valuable \emph{memory layers}: they orchestrate extraction,
summarization, and retrieval over general-purpose backends. We position
FluctlightDB as the missing \emph{engine} beneath such layers---where the native
operations are \texttt{experience} and \texttt{activate}, not SQL or
\texttt{vector\_search}.

\subsection{Positioning}
We propose an engine-level data model for long-term agent memory---episodes, cues,
provenance on write; activation on read---distinct from SQL rows and vector ANN.
We make this case carefully, not categorically: Mem0~\cite{chhikara2025mem0},
Mem0$^{\text{\tiny g}}$, Zep~\cite{zep2024}, and HippoRAG~\cite{gutierrez2024hipporag}
already treat memory as structured facts, graphs, or associative indices with
provenance-like trust policies in application layers. Our claim is narrower: an
\textbf{embedded} store whose native contract is \texttt{experience}/\texttt{activate},
with durability, separation, and hybrid recall in one binary---a substrate those
layers can sit on, not a replacement for their extraction and orchestration.
Enterprise databases adding agent features (e.g.\ Oracle AI Database unified memory)
retrofit memory onto existing engines rather than defining episodes and cues as
first-class types. Table~\ref{tab:landscape} summarizes the landscape; direct
head-to-head retrieval on LoCoMo evidence recall against Mem0/Zep is future work
(different published metrics today).

\begin{table}[h]
\centering
\caption{Landscape: memory \emph{layers} vs.\ FluctlightDB as an embedded \emph{engine}.
\textbf{Top block:} our internally reproduced metrics. \textbf{Bottom block:} cited
vendor/literature figures on \emph{different} metric types---not head-to-head.}
\label{tab:landscape}
\small
\begin{tabular}{p{2.0cm}p{1.6cm}p{2.4cm}p{2.6cm}p{2.2cm}}
\toprule
\textbf{System} & \textbf{Kind} & \textbf{Native contract} & \textbf{Benchmark} & \textbf{Metric type} \\
\midrule
\multicolumn{5}{l}{\textit{FluctlightDB (this work; internally reproduced)}} \\
\midrule
FluctlightDB & Engine & \texttt{experience} / \texttt{activate} & \textbf{96.8\%} & LoCoMo evidence (raw, $k{=}150$) \\
FluctlightDB & Engine & hybrid index + session keys & \textbf{97.6\%} & LongMemEval session@8 \\
FluctlightDB & Engine & Muon + paper E2E & \textbf{97.4\%} & LongMemEval E2E QA \\
\midrule
\multicolumn{5}{l}{\textit{Cited vendor/literature (different metrics; not comparable)}} \\
\midrule
Mem0 / Mem0$^{\text{\tiny g}}$ & SDK + backend & Extract / consolidate / retrieve facts & $\sim$92\%+ LLM-J & End-to-end QA \\
Zep & Managed layer & Temporal KG + summaries & $\sim$75\% LLM-J & End-to-end QA \\
Cognee & Pipeline & Graph + vector ETL & --- & Task-specific \\
MemGPT / Letta & Agent OS & Context tiers / blocks & --- & Session QA \\
HippoRAG & Graph RAG & Associative graph retrieval & --- & Multi-hop QA \\
Chroma / Qdrant & Vector DB & ANN top-$k$ & --- & IR / similarity \\
\bottomrule
\end{tabular}
\end{table}

Benchmarks LoCoMo~\cite{maharana2024locomo} and LongMemEval~\cite{wu2024longmemeval}
measure long-horizon memory; BEIR~\cite{thakur2021beir} anchors IR credibility.
Our framing---memory as a \emph{proposed} data model on equal footing with rows and
vectors---is the argument this engine and its measurements are built to test.

\section{Conclusion}
The relational model gave applications a database for \emph{facts}; the vector
model gave search a database for \emph{similarity}. Autonomous agents need a
database for \emph{memory}, and it should be as boring to adopt and as rigorous to
trust as SQLite. FluctlightDB is our argument that this engine can exist today: it
ties vector baselines where they are strong on shared harnesses, reaches 96.8\%
raw LoCoMo evidence \emph{retrieval} (no neighbor expansion) with $\approx$85\%
end-to-end QA over date-stamped context, scores 97.6\% session recall and 97.4\% end-to-end QA
on LongMemEval-S, and---on the provenance thesis under shared memory---only 18\%
top-1 on the graded conflict suite (100\% under isolation). Harnesses, frozen
JSON, and MIT engine source are released so others can re-run and contest these
claims; the one-minute path uses the published wheel.

\appendix
\section{Recall Fabric mechanisms}
\label{app:fabric}
Hybrid recall (Section~\ref{sec:design}) treats memory as an index. FluctlightDB also
implements a \emph{Recall Fabric}---Photon prefilter (SimHash/LSH), Manifold Lattice
addressing~\cite{fiete2008,sreenivasan2011}, theta--gamma phase
parsing~\cite{lisman1995,lisman2013,plate1995}, consolidation crystallization,
adaptive forgetting, chronos time buckets, and confidence-weighted
consensus---each as a deterministic, dependency-free module behind
\texttt{FLUCTLIGHT\_FABRIC=1}. On ingest, Fabric indexes engrams (and CHORUS traces)
for lattice/phase scoring; on recall it re-ranks hybrid and CHORUS shortlists before
returning hits. Synthetic property tests cover CRT round-trip, bind/unbind
invertibility, Hamming--cosine rank agreement, and decay monotonicity. Unlike the
headline results in Section~\ref{sec:eval}, these mechanisms are \emph{not}
validated as standalone peer benchmarks---only as optional rerank on CHORUS shortlists
where Fabric was enabled. Extended mechanism notes live in repository documentation.

\begin{figure}[h]
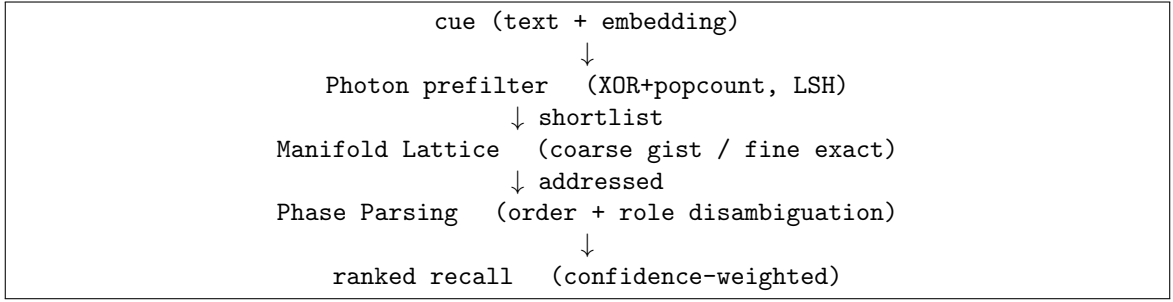

\centering
\fbox{\parbox{0.92\columnwidth}{\centering\ttfamily\small
cue (text + embedding)\\
$\downarrow$\\
Photon prefilter \; (XOR+popcount, LSH) \\
$\downarrow$ shortlist\\
Manifold Lattice \; (coarse gist / fine exact) \\
$\downarrow$ addressed\\
Phase Parsing \; (order + role disambiguation) \\
$\downarrow$\\
\textbf{ranked recall} \; (confidence-weighted)
}}
\caption{Recall Fabric pipeline (\texttt{FLUCTLIGHT\_FABRIC=1}). Composes with CHORUS
recall on LoCoMo, BEIR, and FAMB frozen runs (Section~\ref{sec:eval}).}
\label{fig:fabric}
\end{figure}

\section*{Artifacts}
Repository: FluctlightDB (MIT). Harnesses: \texttt{benchmarks/locomo\_eval.py},
\texttt{benchmarks/locomo\_ablation.py},
\texttt{benchmarks/provenance\_conflict\_bench.py},
\texttt{benchmarks/longmemeval\_bench.py}, \texttt{benchmarks/longmemeval\_colab.ipynb},
\texttt{benchmarks/beir\_bench.py}, \texttt{benchmarks/agent\_memory\_bench.py}.
Frozen metrics: \texttt{benchmarks/results/paper-2026-07-10.json}. E2E:
\texttt{benchmarks/longmemeval\_e2e.py};
\texttt{benchmarks/results/e2e-cert-paper-v2-2026-07-07.json}. Preprint DOI:
\url{https://doi.org/10.5281/zenodo.20949890}. Source:
\url{https://github.com/voxmastery/FluctlightDB/tree/main/papers/arxiv-v1}.

\bibliographystyle{plain}
\bibliography{references}

\begin{thebibliography}{10}

\bibitem{chhikara2025mem0}
Prateek Chhikara, Dev Khant, Saket Aryan, Taranjeet Singh, and Deshraj Yadav.
\newblock Mem0: Building production-ready ai agents with scalable long-term
  memory.
\newblock {\em arXiv preprint arXiv:2504.19413}, 2025.

\bibitem{codd1970relational}
Edgar~F Codd.
\newblock A relational model of data for large shared data banks.
\newblock {\em Communications of the ACM}, 13(6):377--387, 1970.

\bibitem{fiete2008}
Ila~R. Fiete, Yoram Burak, and Ted Brookings.
\newblock What grid cells convey about rat location.
\newblock {\em Journal of Neuroscience}, 28(27):6858--6871, 2008.

\bibitem{gutierrez2024hipporag}
Bernardo Guti{\'e}rrez, Yiheng Shu, Yu~Gu, Peter Pasupat, Barret Liu, Rishabh
  Shah, Ming-Wei Chang, Ramakanth Pasunuru, Yujia Zhang, Avi Gupta, et~al.
\newblock Hipporag: Neurobiologically inspired long-term memory for large
  language models.
\newblock {\em arXiv preprint arXiv:2405.14831}, 2024.

\bibitem{sqlite2024}
Richard~D. Hipp.
\newblock {SQLite}.
\newblock \url{https://www.sqlite.org/}, 2024.
\newblock Embedded relational database engine.

\bibitem{lisman1995}
John~E. Lisman and Marco A.~P. Idiart.
\newblock Storage of $7 \pm 2$ short-term memories in oscillatory subcycles.
\newblock {\em Science}, 267(5203):1512--1515, 1995.

\bibitem{lisman2013}
John~E. Lisman and Ole Jensen.
\newblock The theta-gamma neural code.
\newblock {\em Neuron}, 77(6):1002--1016, 2013.

\bibitem{maharana2024locomo}
Adyasha Maharana, Dong-Ho Lee, Sergey Tulyakov, Mohit Bansal, Francesco
  Barbieri, and Yuwei Fang.
\newblock Evaluating very long-term conversational memory of llm agents.
\newblock In {\em ACL}, 2024.

\bibitem{packer2023memgpt}
Charles Packer, Sarah Wooders, Kevin Lin, Vivian Fang, Shishir~G Patil, Ion
  Stoica, and Joseph~E Gonzalez.
\newblock Memgpt: Towards llms as operating systems.
\newblock {\em arXiv preprint arXiv:2310.08560}, 2023.

\bibitem{plate1995}
Tony~A. Plate.
\newblock Holographic reduced representations.
\newblock {\em IEEE Transactions on Neural Networks}, 6(3):623--641, 1995.

\bibitem{sreenivasan2011}
Sameet Sreenivasan and Ila Fiete.
\newblock Grid cells generate an analog error-correcting code for singularly
  precise neural computation.
\newblock {\em Nature Neuroscience}, 14(10):1330--1337, 2011.

\bibitem{thakur2021beir}
Nandan Thakur, Nils Reimers, Andreas R{\"u}ckl{\'e}, Abhishek Srivastava, and
  Iryna Gurevych.
\newblock Beir: A heterogeneous benchmark for zero-shot evaluation of
  information retrieval models.
\newblock In {\em NeurIPS Datasets and Benchmarks}, 2021.

\bibitem{cognee2024}
{Topoteretes}.
\newblock Cognee: Memory engine for ai agents.
\newblock \url{https://github.com/topoteretes/cognee}, 2024.
\newblock Vector + knowledge-graph memory pipeline.

\bibitem{wu2024longmemeval}
Di~Wu, Hongwei Wang, Wenhao Yu, Yuwei Zhang, Kai-Wei Chang, and Dong Yu.
\newblock Longmemeval: A benchmark for long-term memory in llm chat assistants.
\newblock {\em arXiv preprint arXiv:2410.10813}, 2024.

\bibitem{zep2024}
{Zep AI}.
\newblock Zep: A temporal knowledge graph architecture for agent memory.
\newblock \url{https://www.getzep.com}, 2024.
\newblock Agent memory platform; LoCoMo evaluations in product documentation.

\end{thebibliography}
\end{document}